# Decolonial AI as Disenclosure

**Warmhold Jan Thomas Mollema** 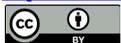

Graduate School of Natural Sciences, Department of Information and Computing Sciences, Utrecht University, Utrecht, Netherlands
Email: wjt.mollema@gmail.com

## Abstract

The development and deployment of machine learning and artificial intelligence (AI) engender "AI colonialism", a term that conceptually overlaps with "data colonialism", as a form of injustice. AI colonialism is in need of decolonization for three reasons. Politically, because it enforces digital capitalism's hegemony. Ecologically, as it negatively impacts the environment and intensifies the extraction of natural resources and consumption of energy. Epistemically, since the social systems within which AI is embedded reinforce Western universalism by imposing Western colonial values on the global South when these manifest in the digital realm is a form of digital capitalism. These reasons require a new conceptualization of AI decolonization. First this paper draws from the historical debates on the concepts of colonialism and decolonization. Secondly it retrieves Achille Mbembe's notion of decolonization as disenclosure to argue that the decolonization of AI will have to be the abolishment of political, ecological and epistemic borders erected and reinforced in the phases of its design, production, development of AI in the West and drawing from the knowledge from the global South. In conclusion, it is discussed how conceiving of decolonial AI as form of disenclosure opens up new ways to think about and intervene in colonial instantiations of AI development and deployment, in order to empower "the wretched of AI", re-ecologise the unsustainable ecologies AI depends on and to counter the colonial power structures unreflective AI deployment risks to reinforce.

## Keywords

AI, Colonialism, Decolonization, Disenclosure, Mbembe

## 1. Introduction

Algorithmic digital technologies, machine learning (ML), and other forms of Artificial Intelligence (AI) that rely on huge datasets are currently central to digital capitalism (Ricaurte, 2022: p. 727). The constituents of digital capitalism—cloud computing enabling data centers and the generation of energy they require, the mining of raw materials and the making of data—make these technologies possible, all through forms of extraction and labor (Jung, 2023). The world's most powerful companies and states reinforce themselves through the usage and development of AI technologies. Consequently, the operation of those technologies demands more resources every day and new forms of colonization revolving around AI, that Achille Mbembe calls "techno-molecular colonialism" have emerged (Mbembe, 2022: p. 32). This is the fusion of colonialism's exploitative and extractive tendencies with capitalization that extends to the "molecular" depths of human behavior and the Earth's ecologies. Some consider digital capitalism[1] synonymous with "data colonialism" in the way data is extracted from life forms' quantifiable traces. Especially the global South[2] is at risk, where digital protections are less secure (Couldry & Mejias, 2019: xix-xx)[3]. At the root of the techno-molecular colonialism lies the magnetic power of the "techno-colonial market" scales from the neuronal level of behavioral predictions up to the global level dynamic of resource extraction, treating everything as "raw material" ready for datafication. The key features of this new form of marketization are that it is—technologically speaking—an algorithmic structure enabling cycles of data extraction from quantifications of life forms (Mbembe, 2022: p. 67) that—economically speaking—capitalize upon datafication in a way that "reinvigorate[s] and rework[s] colonial relationships of dependency" (Madianou, 2019: p. 2). As such it constitutes a paradoxical network of power that is capable of extracting wealth from catastrophes it has prefigured itself, rather than balking at their imminent occurrence (Mbembe, 2022: p. 3).

For the purposes of this paper, the coalescence of digital capitalism's drive for the extraction of natural resources and data and Mbembe "techno-molecular colonialism" are subsumed under the term "AI colonialism" because of the colonial characteristics they share. Decolonial approaches to data, AI, and technology connect the consequences of digital capitalism to the persistence of his-torical colonialism's harms. As an explanatory mechanism this imports the structural injustices of the already centuries-long tale of domination that is colonialism (Butt, 2013). Placing AI injustices under the banner of AI colonialism, in continuity with historical colonialism, therefore leads to a better understanding of the specific contemporary ideologies of dataism, ethics washing, and digital domination, while at the same time approaching them as structural injustices (Couldry & Mejias, 2019: p. 11)[4].

---

[1]It still has function to speak of "capitalism" without an adjective, as Couldry and Mejias note, because even though the adjective "digital" is very characteristic of the frontier of capitalistic development and the corporations that dominate the global economy, "digital capitalism" is by no means decoupled from the general system of the "organization of life so as to maximize surplus value, resulting in the concentration of power in very few hands". However, I will continue to use the adjective "digital" because it does manage to draw attention to the new dimensions AI technologies introduce to the trajectory of capitalism (Couldry & Mejias, 2019: ch. 1, p. 18).

[2]By "global South", I understand the cross-border non-hegemonic "plural entity" of populations of the Earth. Geographically, the Global South is mostly comprised out of African, Asian and South American peoples, but within the bastions of the Global North, oppressed and marginalized "Souths" exist as well. Next to that, the North is not only present in Europe, North America, China and Russia; the North also exists in the oppressive elite and allied structures that occupy the world's Souths (Arun, 2020: p. 593). In the intentionally manifestoed words of Boaventura de Sousa Santos "We are the global South, that large set of creations and creatures that has been sacrificed to the infinite voracity of capitalism, colonialism, patriarchy." (Santos, 2014: p. 16). A useful synonym for "global South" used in this sense is "majority world" (Ricaurte, 2022).

[3]Couldry and Mejias define "data colonialism" as "*the mutual implication of human life and digital technology for capitalism*," in which social life is directly appropriated as economic factor of production (Couldry & Mejias, 2019: ch. 1, 5-6).

Correspondingly, the central question this paper is occupied with, taking colonial AI injustices as a target, is "how should the decolonization of AI be conceptualized?" The answer defended here is that decolonial AI should be conceived of as the abolishment of political, ecological, and epistemic borders that AI technologies erect or reinforce in the phases of design, production, and development. To corroborate this answer, two subservient aims will be realized. Firstly, three claims for why AI needs to be decolonized are presented. Secondly, multiple anticolonial critiques of AI are discussed, and it is explored how they could be synthesized into a general conception of decolonial AI that adequately deals with its political, epistemic and ecological facets.

## 2. Methodology and Structure

The philosophical methodology employed in this paper is (i) to construct a conceptual framework that functions as a basis for (ii) the examination of a contemporary phenomenon. Next, (iii) related philosophical work is represented in the function of case studies. At the end of the paper, (iv) the resulting insights be discussed.

(i) Qua conceptual framework, the concept of decolonialism is reconstructed and the terms "decolonization", "decolonialism", or "decoloniality" are clarified (section 3). (ii) Using the conceptual framework of decolonialism, AI colonialism is examined and political, ecological and epistemic reasons for why the decolonization of AI has to be undertaken are stipulated (section 4). This results in the definition of a new approach to decolonial AI: "decolonial AI as disenclosure" that is further developed using Achille Mbembe's conception of decolonization as disenclosure as a general framework for the decolonial discourse (section 5). (iii) Subsequently, topical studies of approaches to decolonizing AI that centralize different concepts and values—such as precaritisation, centralization of power, sociotechnical foresight, participatory AI and relational autonomy—are presented (section 6). Finally, this paper will end with a discussion of decolonial AI as disenclosure as related to the political, ecological and epistemic reasons for decolonizing AI (section 7). It is concluded that decolonizing AI should amount to disenclosing the colonial forms of oppression and domination AI makes possible.

## 3. What Is Decolonization?

To understand the meaning of the term decolonization, approaches from the

---

[4]For a general discussion of structural injustice, resort to (Young, 1990).

African continent as well as from the Americas that define what colonialism is and how to undo its effects are retrieved.

Canonically defined, colonialism is a conglomerate of "…*domination*, *cultural imposition* and *exploitation*" (Butt, 2013). Following Iris Marion Young, the difference between "domination" and "oppression" is that domination is the obstruction of a social groups self-determination, while oppression is the inhibition of social group's expressing and doing (Young, 1990). Historically, the phenomenon can also be decomposed into four other components: appropriation of resources; inequal social relations that fortify appropriation; inequality in the distribution of the benefits from the economic processing of resources; and the spread of dominating ideologies that reframe and justify this situation (e.g. via racism) (Couldry & Mejias, 2019: p. 10). It can therefore be recognized that colonialism has both material and cognitive/epistemic dimensions. In short, colonialism is a historical and actual pattern of injustice that Western potentates in South America, Asia, and Africa weaved (Getachew, 2019).

In the late nineteenth century for South America and in the mid-twentieth century in the African context, "decolonization", as historian Adom Getachew explains, was understood as the political regime change from the external rule of imperialist nations over their colonies towards the emergence of postcolonial societies and nation-states (Getachew, 2019: ch. 2-3). After the right to self-determination was inscribed in the United Nations Charter, for the African states that emerged in 1950-60 in particular, such as Ghana, and Guinea-Bissau, decolonization meant moving away from Western domination towards a nationalist nation-state that sought to empower natives and their culture. However, as Getachew argues, the term "decolonization" was quickly appropriated by Western potentates, who framed it—in line with racist hierarchies in the League of Nations and the Eurocentrism inscribed into the UN's origin and outlook—as a "natural consequence" of empire and a succession into civilization that enabled imperialist nations to trap the emerging nations into structures of economic dependency that safeguarded the future of Western capitalism in the global South (Getachew, 2019: ch. 2). It is therefore clear that "decolonization" is not an uncontested term.

To come to terms with an unappropriated and disruptive sense of the concept, anti-colonial and decolonial scholars whose thoughts are uncolored by capitalist agendas are consulted. One such thinker is the Cameroonian philosopher Achille Mbembe, who explains on multiple occasions that a starting point to think about decolonializing is to conceive of it as a multifaceted *summons*, a summons for changing the dominating relations of power that are characteristic of colonialism (Mbembe, 2015).

Colonialism, as we have defined it, has already found its way as an explanatory concept and goal to abolish into scholarly work engaged with digital technologies. South African scholar Rachel Adams includes the work African philosophers and the work of South American scholars of the "decoloniality" variant of colonialism, such as the Peruvian scholar Aníbal Quijano, in her analyses of AI (Adams, 2021). Likewise, the Mexican scholar Paola Ricaurte builds upon Quijano's concept of the "coloniality of power" to extend it to functioning as an analytical model of how coloniality as an overarching concept explains the pervasion of domains like the economy, politics, being/sensing, nature, and technology by the extractive rationality of digital capitalism (Ricaurte, 2019: pp. 354-357). Translating this into the analysis of the digital global economy, as Couldry & Mejias have done, results in recognizing that several currents and movements of critique of this type of capitalism, such as those emphasizing either the racial, heteronormative or colonial aspects of capitalism, centralize various overarching premises that coalesce in their "logics". Because of this coalescence, these reactions can be grouped under the umbrella of the "decolonial turn" that is happening in data and technology studies (Couldry & Mejias, 2019: p. 8). What these scholars all recognize is that the revolutionary aspect of decolonization cannot be emphasized enough. Calls to actively rethink the deep influence of oppressive colonial modes of thought on the being of social groups often risk being co-opted by reformist causes, which takes away the radical thrust of decolonial theory—which will be dealt with in more detail in section 7.1. (Adams, 2021: p. 2, 7). This is a contemporary pitfall that decolonial theory faces today, reminiscent of the imperialistic appropriation of decolonization Getachew identified (Getachew, 2019: ch. 7).

Backed by this understanding of colonialism, the concept of *de*colonialism can now be discussed. One of the cornerstone figures of the theoretical movement of decolonization is the Kenyan writer Ngugi wa Thiong'o, who helped popularize the intellectual project of decolonization in the 1980s. In his work, decolonization is a project of *re-centering* the mind of colonized peoples on their own culture, language, and self-worth through forms of education (Mbembe, 2015). This conception emphasises a cognitive and epistemological component of colonialism that must be abolished, which underlies the economic and political variant of decolonization that focuses on banishing alien rule.

In the work of Frantz Fanon, writing 20 years before Ngugi wa Thiong'o, decolonization appears as a retainment of self-ownership that frees the colonized from "the gap between image and essence" (Fanon, 2008: p. 168). Fanon identified the cognitive, epistemological, and phenomenological barriers imposed on the colonized by colonialism. By the word "gap" he means the rift between, on the one hand, how the colonized are defined externally based on their skin and, on the other hand, the difference of their own becoming, free from domination. Central in Fanon's work for theorizing decolonization was the idea of the "*resurgence of man*" from his colonized, objectified, state into a "becoming-human" that is not that of the colonizer, but fully the colonized's own. Fanon characterized the colonial experience as expressing a "double consciousness": self-consciousness and the additional consciousness of oneself through the racializing gaze of the Other (the colonizer). Double consciousness leads to the colonized embodying and internalizing the ontological opposition between the being of a Black human and the white oppressor (Fanon, 2008: pp. 169-170). Fanon formulated it radically when he explained that colonial whiteness, as a culturally dominating form of life, *epidermalizes* the Black man's degradation from man into object. The neologism "epidermalizes" denotes how colonial domination fixates the degradation and dehumanization of the Black man to the domain of the subhuman onto the Black man's own skin (Fanon, 2008: p. 172).

The grafting of cognitive and epistemic complexes onto colonial subjects is backed up by historical assertions such as that of the African scholar Ndlovu-Gatsheni, who for instance writes that in colonialism, "the commercial non-territorial empire and the cognitive empire are, inextricably intertwined" (Ndlovu-Gatsheni, 2019: p. 208). Similarly, the contemporary Nigerian philosopher Uchenna Okeja describes colonization as a "conceptual adjustment program" that replaced the rule of customs by written law, polygamy by monogamy, and polytheism by monotheism through the exertion of extreme forces of domination and conversion, thereby committing an epistemic vice (Okeja, 2022: p. 205). Like Fanon, modern decolonial scholar Santos also stresses that the duality "appropriation/violence" is at the heart of colonialism. For Santos, colonialism "can only be tackled if we situate our epistemological perspective on the social experience of the other side of the line, the nonimperial global South, conceived of as a metaphor for the systemic and unjust human suffering caused by global capitalism and colonialism" (Santos, 2014: p. 208).

To sum up, colonialism has preyed on peoples' lands, knowledges, customs, and consciousness. Decolonization, the negation of colonialism, dictates that future generations in the postcolonial societies that emerged in the late nineteenth and mid-twentieth centuries, and societies in the "postimperial" West, have a collective responsibility to revisit, if not undo, the forced conceptual adjustments, states of consciousness and economic dependencies inherited from colonialism (Ndlovu-Gatsheni, 2013). In the context of Latin American independence, Quijano has described the decolonial imperative in similar terms as "to cease being what we are not." He viewed the freeing of oneself from superimposed images of Eurocentrism as a starting point for redistributions of power in postcolonial societies (Quijano & Ennis, 2000: p. 574)[5].

What the currents of decolonization mentioned up to now have in common is that decolonial resistance has as one of its goals to change the colonizeds' relation to themselves. One can see how this relation is instantiated in colonialism and can be addressed if Mbembe's explanation of the concept of *racialization* as inherent to the practice of colonialism is followed: "Racialisation must be understood as the capture and conscious deployment of a set of techniques of power […] that aim at producing a reality, namely race, that there is then a concerted attempt to naturalize" (Mbembe, 2022: p. 15). Race was constructed to function as a barrier between citizenship and identity (Mbembe, 2019: p. 61).

[5]For Quijano, the coloniality of power controls four aspects: labor (capitalism), sex (bourgeois family), authority (nation-state) and intersubjectivity (Eurocentrism).

The lasting implications for the (formerly) colonized's sense of identity were grave, consistent with what Fanon has shown. The imposition of the racial as defining aspect of the boundary between the human and the subhuman has had lasting implications for the social ontologies underlying Western societies' social contract as well, and, with the late Charles Mills, one can argue that "whiteness"—not qua color but qua ontological category—has become the defining property of who counts as human in these societies (Mills, 1998).

Synthesizing the positions of Fanon, wa Thiong'o and Okeja on the topic, decolonization can hence be understood as the space of intersections of actions that form, simultaneously, (a) a resurgence of identity and political self-determination in the reaffirmation of a denied difference and (b) a return to culture and language for conceptual readjustment to deny the naturalizing move of imposed racial categories.

## 4. Why is AI in Need of Decolonization?

Having grasped (i) *colonialism* to be the historical and actual extractivist pursuit of natural resources and cultural and epistemic dominion, and (ii) *decolonization* to be the multifaceted realization of colonialism's negation in action, concepts like "data colonialism" and "AI colonialism" can be analyzed. The term "data colonialism" signifies how in our digital age, new types of resource appropriation, unequal relations and distributions are grafted upon and in continuity with existent components of colonialism (Couldry & Mejias, 2019: pp. 11-14). AI colonialism conceptually overlaps with data colonialism, as the data extraction the latter emphasises is in service of the former. But they are inequivalent, since the usage of the term "AI colonialism" wants to stress the way in which injustices (forms of domination and oppression) characteristic of the colonial are perpetuated, amplified and instigated in the development and deployment of AI systems. These AI systems are technologies of the automatization of rudimentary to proficient levels of human capacities to which responsibilities are outsourced or which enable new forms of human interaction. The intertwinement and continuity of digital capitalism with historical colonialism and their dependency on material ecosystems, make today's digital capitalism and AI colonialism similar to and intensifications of their historical predecessors. "[G]lobal capitalism has never been so avid for natural resources as today, to the extent that it is legitimate to speak of a new extractivist imperialism," says Santos (2014: p. 43) and with the advent of datafication, AI colonialism is avid for the extraction of new resources. However, for datafication to succeed, human life must first be controlled and reconfigured to become "ripe for the picking" (Couldry & Mejias, 2019: p. 5). Likewise, Ricaurte discerns that "Massive unidirectional data flows from south to north contribute to increasing wealth concentration in a few industrialised countries and their companies, the production of poverty by expanding dispossession, resulting in technological and epistemic dependence and global epistemic injustice" (Ricaurte, 2022: p. 731). Derivatively, the summons decolonization of AI represents, is to counter these forms of colonization inherent in and exhibited by AI systems. Anticolonial action needs to be undertaken at macro-, meso- and micro-levels of the AI ecology (Ricaurte, 2022: p. 737).

The remainder of the paper will be concerned with the following three reasons for decolonizing AI.

Firstly, the *political* reason for the decolonization of AI is that AI must be decolonized to counter its role in strengthening the dominion of digital capitalism. Digital capitalism's course is shifted by the way intelligence technologies are developed. The role of digital technologies in modern societies leads them to be increasingly convolved by the same concept: that of the computational. In the words of Mbembe, "The computational is generally understood as a technical system whose function is to capture, extract, and automatically process data that must be identified, selected, sorted, classified, recombined, codified, and activated" (Mbembe, 2020: p. 19). Because of the introduction of LLMs, the manufacturing of significantly more efficient neural networks (Kozlov & Biever, 2023) and ever-faster computer chips to speed up AI (Castelvecchi, 2023), anything that can be datafied can also be capitalized upon with a staggeringly inequitable speed. Far from being politically neutral developments, the sped-up AI development risks contributing to economic marginalization and worker exploitation, while enabling injustices with respect to social representation in the digital public sphere, information processing and the commodification of citizens as data resources (McQuillan, 2022). This economic and political imperialism in the digital economy driven by AI technologies is rather a continuity of past injustices.

To summarize the political reason: AI catalyzes the extractivist tendencies of digital capitalism's global economy, extends the historical colonial project and is hence in need of political decolonization.

Secondly, from an *ecological* perspective, it can be argued that AI must be decolonized because of the environmental burden AI colonialism places on planetary natural resources. AI's backbone, the rising data economy, has an ecological footprint with two negative sides to it. 1) It takes exorbitant amounts of electricity and resources to power data centers that are required for this economy's infrastructure (Crawford & Joler, 2018; Dhar, 2020). The mining of minerals like lithium, tin, cobalt, manganese, and nickel needed for the fabrication of data technologies is ravishing and polluting all around the world; the local ecological impacts of the data centers' toxic waste are devastating (Mbembe, 2022: p. 42). These processes are gradually diminishing the elasticity and resilience of ecosystems and species, which are often already in a state beyond repair, so that all that rests us is memorizing their existence as "speakers for the Dead", to use Donna Haraway's term (Haraway, 2016: p. 164). This leads to the asymmetry of Southern countries having to bear ecological disruptions while Northern countries reap the economic benefits of AI. 2) Techno-solutionist narratives accompany the development and deployment of AI. However, the narrative of framing of AI as a force for social good and solution climate change (as DeepMind's founder Denis Hassabis has done (Cave, 2020)) relaxes the imperative to take climate change—or what Bruno Latour called "the Terrestrial as an actor" (Latour, 2018)—seriously. *In extremis*, AI techno-solutionism even induces fantasies of space colonization amongst Big Tech's elites (Crawford, 2021: pp. 234-235).

In recapitulation of the ecological reason: AI requires ecological decolonization because: (a) it deepens schemes of resource extraction that either replicate or intensify colonial geopolitical relationships (the global South being the global North's mining site); and (b) the narrative intertwinement of AI with climate and environment is understudied, which covers up further appropriations.

Thirdly, the *epistemic* reason for decolonizing AI holds that AI is complicit in forms epistemicide by enforcing a false universalism in line with colonial value impositions. Scholars are increasingly calling attention to how AI is enveloped in the production of quasi-universal deceptions (Katz, 2020). AI and ML models have been characterized warningly as "stochastic parrots" (Bender et al., 2021) and "refracturing mirror of the world" (Pasquinelli & Joler, 2021). These are characterizations that revolve around the insight that AI salvages the past and projects it upon the future, thereby importing the injustices of the past. Several scholars have shown how this reifies the injustices that are co-constitutive of the data-histories ingrained into the models' knowledge reproduction. And how this entails that AI cannot but reproduce existing hateful, racist, sexist, homophobic content. Stereotypes like this are deeply embedded into the colonial past of Western society (Davis, 1981). This results in the presentation of this content as "the real", while simultaneously hiding this constructed nature (McQuillan, 2023a). So the problems are *social* rather than technical problems that could be solved by technological band-aids like "correcting for biases" or "explaining black boxes". Such technofixes distract from the gravitas of the cognitive and affective sides of the problem, because these artificially constructed realities are capable of changing people's beliefs as Abeba Birhane has recently argued (Birhane, 2023). Yuval Noah Harari has stressed that the emulation of human empathy and knowledge production AI exhibits, is capable of inciting social disruption by tapping into human civilization's weaknesses (Harari, 2023). As the project of micromanaging global beliefs via explanation of algorithms and bias corrections seems foolish from the get-go, a different response is needed.

Thus, the epistemic reason is: AI systems are in need of epistemic decolonization because of their capacity for reifying Western knowledge and values as universal and the viral reproduction of the colonialist superstructure's cognitive content.

I will return to the political, ecological and epistemic reasons for decolonizing AI in section 6, where they are corroborated by way of topical studies in contemporaneous work on decolonial AI.

It is hypothesized that conceiving of decolonial AI using the concept of disenclosure will satisfy the political, ecological and epistemic reasons for decoloniza-tion. That is to say: via conceptualizing the decolonization of AI systems' development and deployment as a form of *disenclosing* of the barriers and borders these processes institute and reinforce/enforce. This conception of decolonization depends on the drastic political, ecological and epistemic reform of societies. A scholar like Ricaurte is optimistic regarding this endeavor: "We can reverse extractive technologies and dominant data epistemologies in favor of social justice, the defense of human rights and the rights of nature" (Ricaurte, 2019: p. 361). More pessimistically, an author like Crawford quotes Audre Lorde that "the master's tools will never dismantle the master's house," arguing hopes of democratizing AI should be tampered. This is based upon her conclusion that "the infrastructures and forms of power that enable and are enabled by AI skew strongly toward the centralization of control" (Crawford, 2021: p. 223). Accordingly, shifting the focus from embedding ethics into AI to the focus on how AI is embedded in relations of *power* is needed (D'Ignazio & Klein, 2020: ch. 1); this paper can be situated within that context. Whether in the optimistic or pessimistic camp with respect to decolonial AI as disenclosure, both camps agree AI injustices are always entangled with other forms of injustice, and that "[t]he calls for labor, climate, and data justice are at their most powerful when they are united" (Crawford, 2021: pp. 226-227).

## 5. Decolonization as Disenclosure

To proceed, the notion of "decolonization as disenclosure" is elaborated upon, based on Achille Mbembe"s account of decolonization. Mbembe centralizes the concept of "disenclosure"—borrowing the term from the work of Jean Luc Nancy. "Decolonization as disenclosure" builds directly upon the thought of Fanon that was surveyed earlier as one of the only real *theories* of decolonization. Centralizing the concept of disenclosure means principally that decolonization amounts to the bringing down of borders that have been erected by colonial domination and oppression in land, body, and mind (Mbembe, 2019: p. 53).

With regard to this project, some preliminary remarks are in order. The subsumption is not uncontroversial because some scholars have conceived of colonialism as something *episodic* (Ndlovu-Gatsheni, 2019: p. 210). But colonialism has not been abolished anywhere but has been a historical continuity since the sixteenth century. In the last century, the global North has also turned colonialism in on itself geographically, turning parts of the global North into "Souths" to be colonized (Couldry & Mejias, 2019; Santos, 2014). Also, for those hesitant to use "coloniality" and "colonialism" side by side—as the former should denote the *survival* of the power structures attributable to the latter's *legacy*—it can be contested if the latter has disappeared at all. Rather, one should be cognizant of the differences in colonialism's persistence in *regional* contexts (say, Sub-Saharan Africa vs. South America vs. Southeast Asia). In short, I agree with Ndlovu-Gatsheni when he says "[…] it would seem that coloniality and colonialism tend to refer to the same situation" (Ndlovu-Gatsheni, 2019: p. 213). Both con-ceptualizations of the residue of the colonial take race as the central organizational principle. Equally, all movements in decolonial thought discussed up until now can be viewed as species of *reparative justice*, i.e. they are concerned with restoration, reconciliation, renarration and other forms of undoing and addressing harms that have a colonial origin.

Given this preliminary framing, Mbembe's view can be discussed in detail. Mbembe reads two layers into Fanon's decolonial thought in order for him to argue for decolonization as disenclosure. The first is that of the triumph of difference over the repetition of racialization and the restitution of the colonized's self to their image The second is the resurgence (*faire surgir*/*surgissement*) of a new "species of man" (*species* in the sense of: appearance) towards the full humanity that was denied in the colonial status quo into self-ownership (Fanon, 2008; Mbembe, 2019: pp. 53-55, 62). Both this resurgence and this restitution (as conceptual children of reparation) are a *taking away of boundaries* within oneself and in one's relation towards the world. Hence they can be typified as *dis*enclosure, the abolishment of oppressive barriers. As we saw in the explanation of the role of racialization in colonialism, "race itself" is such an "enclosure of the world" (Mbembe, 2019: pp. 62-63). With decolonization as disenclosure being capable of incorporating the Fanonian/Quijanoan resurgence/ceasing-what-one-is-not, it becomes clear why Mbembe holds that the "colonial" itself can be defined generally as complete refusal of togetherness and the enclosure of identity and land and the commodification of freedom (Mbembe, 2022: p. 110).

Disenclosure is, therefore, fundamentally a movement of "opening up of the self" towards an unrestricted world of humanity (Mbembe, 2019: p. 63). It is a reparation of the current dichotomy between several "species of man" into a shared horizontal world, one of mutuality and conviviality, as Mbembe reads in the writings of Édouard Glissant and Paul Gilroy (ibid., p. 63). The scale of the decolonial project is planetary after all: the restoration of "*le Tout-Monde*" or *le totalité monde* ("the All-World") and the totality of relations, as Glissant calls it (Glissant, 1997: p. 27), to the resurgence of man that Fanon spoke of. The insight that colonialism is a planetary experience of oppression, takes us back to the realization that colonialism is part of the *hyperobject* that is global capitalism (to use one of Timothy Morton's terms for objects on a grand scale in their dimensional spreading over time and space and capacity to dominate relations with other objects (Morton, 2010)).

Mbembe asserts the postcolonial theory that decolonization is part of, requires thinking of entanglement and concatenation. Proper decolonization, he argues, "insists that identity originates in multiplicity and dispersion, that reference to oneself is only possible within the *between-two*, at the interstice between *marking and unmarking*, in *coconstitution.*" (Mbembe, 2019: p. 73). This unmasks the universalism carried and imposed by colonialism, while at the same time showing how colonialism became a planetary experience that impossibilized any form of coconstitution of difference in the "between-two", as one of those two, being colonized and Othered, was always reduced to the identity of the oppressor (the "subject" of the two). Colonialism, on a cognitive and epistemic level, was always the death of the non-western part of the conversation; what Santos calls "the epistemicide of the South" (Santos, 2014).

The "difference" Mbembe talks about must explicitly not be understood as "an act of disconnection and separation". Instead, it needs to be made clear that decolonial acts, if they are to be, to use my own words, *reparative* rather than segregative, "a set of continuous, entangled folds of the whole" (Mbembe, 2019: p. 80). This decolonial critique of universalism's obstruction of a celebration of difference requires that development and deployment of AI must be reflective of the "…general epistemology of the impossibility of a general epistemology" (Santos, 2014: p. 302). Currently, however, AI contributes to the positioning of Western scientific and predictive knowledges as *universal* upon other ecologies of knowledges, as colonialism has done historically in a very eradicative fashion (Muldoon & Wu, 2023).

Mbembe enquires into planetary entanglement to discern what might be the last utopia imaginable: a new vision of the Earth. This new vision is a *commonisation* of the rights to breathe and sojourn on the planet Earth. Fundamentally, he argues, our "relationship to the Earth is that of the *passer-by*. The Earth receives us and shelters us as passers-by." It is the source of all life, but contrarily to how it is appropriated, mined, and exploited as a tool, the Earth is not something we can take with us. Death, after all, means *returning to the soil*, an insight which leads Mbembe to claim that there is "consubstantiality between the soil and the human person" (Mbembe, 2022: p. 25). Likewise, the metaphors of "fabric" or "weave" that Mbembe employs in the context of postcolonial thought are important and have ecological significance: the metaphors help to think about the interconnection of the world and the planetary scale on which a megaprocess like colonialism takes place. Both weave and fabric point to the stringing together of both objects and species into the world that encompasses the violent tapestry of colonialism (Mbembe, 2022: pp. 86-93). I dwell on these images Mbembe uses because they are particularly congruent with Donna Haraway's slogan that "it matters what thoughts think thoughts," "what systems systematize systems" and "what stories tell stories" (Haraway, 2016: p. 12). The epistemic/cognitive axis of decolonization shows that "it matters what *images imagine images*".

In opposition to the need for commonisation identified by Mbembe stand the tendencies of borderisation by forms of computational/algorithmic "necropolitics". This is the transferal of the political decision-making regarding letting live and letting die to computational/algorithmic systems (Mbembe, 2022). Essentially, necropolitics continue to colonize the global South's resources and people (and those of the Souths in the global North) and try to reinforce a recentralization of a declining Western model of life forms. Mbembe's term necropolitics has also been applied to AI. Ricaurte uses it to describe the regulation of life through AI when she writes: "AI serves as a biopolitical as well as a necropoliti-cal mechanism for regulating life and death" (Ricaurte, 2022: p. 728). Likewise, McQuillan employs the notion as lens to analyze a "fascist politics" as driving AI development and deployment (McQuillan, 2022: p. 97). This is in line with the rise in social fascism through the collapse people's expectations and leads to the othering and exclusion of social groups that is identified by a decolonial scholar like Santos (2014: pp. 84-85). AI can hence be a seen as a technology capable of working in reinforcement of "politics of enclosure".

Politics of enclosure are not new to the history of colonialism. For example, as Toby Green describes in a piece of journalism discussing the significance of the revolutions that have swept Africa in the past five years, in Africa, "the creditor is in power" (Green, 2023). Following the decolonization movements of the twentieth century, the global North has underdeveloped the global South into a state of dependency, where decennia of loans and the import of expertise and technology form a chokehold that continues until this day, effectively in continuity with colonial forms of profiting from the periphery (Green, 2023). Today, its haunting features have shapeshifted into the guise of data and intelligence technologies over the past twenty years (Birhane, 2020). This status quo provides the basis for the borderizations characteristic of colonialism Mbembe talks about and takes us back to the cutting-edge intelligence technologies like biometric identification, facial recognition, location tracking etc., that are deployed to restrain movement and to raise physical and digital barriers of enclosure. These barriers are racialized or otherwise discriminatory in their movement reducing and self-determination impeding nature (Mbembe, 2022: p. 71).

To conclude, the exposition of decolonization of disenclosure yielded a view of decolonization as the *resistance* to seeing the world (i) *from only one perspective* and (ii) *as new, for the taking*—both being perspectival legacies of the Eurocentric colonizer. This resistance must accomplish the constitution of "look[ing] together and eventually to see, […] from several worlds at a time" (Mbembe, 2022: p. 109). Understood in the sense of imposing forms of enclosure on a planetary weave, the way in which the extractivist development, deployment, and production of AI systems operate at the interconnection of colonialism and digital capitalism continues to reproduce and catalyze colonial injustices is a clear demand for the exploration of the conceptual application of the decolonial thought as disenclosure to AI (Joler, 2020).

Having come to terms with decolonialism and conceiving of decolonization in general as disenclosure, the next section deeper explores what it means to apply this to AI.

## 6. Topical Studies: Four Currents in Decolonial AI

Decolonial AI and political philosophy of AI centralizing power over ethics are developing fields (Zimmerman, Vredenbugh, & Lazar, 2022; Birhane et al., 2022a). This section is comprised of four topical studies (exemplary for decolonial technology studies) of how decolonizing AI is conceived of in the litera-ture. The takeaway points of each study are related to the three reasons for decolonization presented in section 4. These four "currents in decolonial AI" are modelled on the "epistemic" types of harmful AI processes Ricaurte has identified: "datafication (extraction and dispossession), algorithmization (mediation and governmentality) and automation (violence, inequality and displacement of responsibility)" (Ricaurte, 2022: p. 727).

Section 6.1. discusses arguments concerning critical cartographies of AI systems. In 6.2., arguments exposing AI's false universalism are discussed. 6.3. looks at resistance to precaritisation and violence instigated by AI. Lastly, 6.4. discusses how the participation of affected social groups can be viewed as a bottom-up method for decolonial AI.

## 6.1. Critical Cartographies of AI Systems

Critical cartography is the tracing the interdependencies of social, technological, economic and political systems. Applied to AI, it describes the complexity of AI systems' dependency on human and natural resources by tracing AI systems' entanglement with colonialism, digital capitalism, and ecocide. An example is the reconstruction of the system Amazon's Echo (predecessor of Amazon Alexa) depends on that Crawford and Joler's undertook. They analyzed and pictured the network and supply chain that is required for Echo's design, production, shipment and functioning. The cartographical exercise's detail shows a sleek design hides an intricate colonial system of mineral extraction, worker exploitation for data labelling, unequal wage distributions, unwarranted data extraction from civilian customers, chaining of customer/Internet/data center/neural network infrastructures and pollution via electronic waste (Crawford & Joler, 2018)[6]. Critical cartography construes the contours of AI systems that make new interventions possible.

Critical cartography corroborates the ecological reason for decolonization because it shows AI systems "are physical infrastructures that are reshaping the Earth, while simultaneously shifting how the world is seen and understood" (Crawford, 2021: p. 28). Crawford identifies "a capture of the commons," where biodata of all varieties are being captured to fuel AI systems' predictions and assessments. Underlying this stance vis-à-vis life are the incessantly intensifying data accumulation and extraction cycles of *datafication* (Crawford, 2021: pp. 114-115). This results in the creation of digital systems that discriminate (classify, order, predict, determine, etc.) on seemingly neutral grounds ("the data speak for themselves"), while in effect, ecosystems and social groups are being subjected to dominative measures that amplify social inequalities (Crawford, 2021: p. 119).

To sum up, critical cartographies of AI systems create decolonial metaphors and insights because of their description of AI systems' ecological effects and economic dependencies. This corroborates the ecological reason for decolonizing AI because the ecological effects and natural and economic dependencies of AI systems come to the fore, which supports the ecological reason's two components that (a) AI systems deepen schemes of resource extraction that are intensifications of colonialism and (b) the mystification and obscuration of AI's dependencies masks actual environmental impacts, prohibiting these downsides from slowing its economic momentum.

---

[6]To explain this data extraction and human worker commodification, Crawford and Joler utilize the figure of a triangle to represent it as a Marxian dialectic, which then fractals into itself to adequately represent scale.

## 6.2. Exposing AI's False Universalism

Arguments exposing AI's false universalism harness critical race theory, feminist ethics and intercultural translation (Santos, 2014: ch. 8) to emphasize injustices associated with ML and other technologies' masquerade as a universal, scientific and unbiased "view from nowhere"[7]. As a corollary, it is shown how the argument that AI systems impose Eurocentric and universalistic knowledge and values, marginalizing social groups in the process, supports the epistemic reason for decolonizing AI.

Yarden Katz argues an amalgamation of white universalism, militarism and racial bias has been foundational for AI Historically, he shows, AI was first mainly developed for military reasons, its directions for development integrating seamlessly into the American military-industrial complex. ML was not considered strictly AI, because of the opinions of theorists like Simon and Newell, who received military funding and functioned as spearheads for the field. After Big Data's perdition following whistleblowing of the misuse and malpractice of data processing and collection by corporations and states alike, Big Data processing algorithms were also subsumed under the term AI. AI thus turned out to be a conveniently capacious term under which projects of data-driven extractivism and technological capitalization could be subsumed. Katz concludes the charade of a "view from nowhere", that is actually an exclusionary Western white male's perspective, has been central to AI's development and deployment since its inception (Katz, 2020). After the turn to connectionism in the 1980's and the neural network revival of the 2000's powered up, AI began to usurp the technological construction of reality via a new image: that of "the facts speaking through the data" (Hao, 2020; McQuillan, 2022: p. 50). The rise of search engines, smartphone technology, and social media reinforced this enlargement of the net of data-extraction.

Given this historical entanglement of AI and Western universalism, Mhlambi and Tiribelli resort to "relational autonomy" and the value of Ubuntu in their effort to conceptualize decolonial AI (Mhlambi & Tiribelli, 2023). Criticizing ethics frameworks and AI applications alike because their concept of autonomy central is non-relational and dependent on the legacy of Enlightenment (Mhlambi & Tiribelli, 2023: p. 868), they argue for applying Ubuntu-type relational autonomy to "…re-imagine AI and the entire ecosystem that produces it. […] AI designed to be relational […] would allow individual users and the community to have more say and power over their online experiences" (Mhlambi & Tiribelli, 2023: p. 877). That adjustment highlights AI systems and policies' protectionism of "Western" autonomy and their simultaneous exclusion of anticipating relational "injustices and inequalities" concerning political/cultural backgrounds unrecognized by systems, policies and frameworks because of Western autonomy's myopic individualism that "[ignores] social aspects that constitute an individual and an individual's ability to make decisions" (Mhlambi & Tiribelli, 2023: p. 877).

---

[7]These approaches identify abstract political problems within AI and turn to philosophical work combatting exclusion and racialization, or values from other and/or excluded cultural contexts to resolve those problems. This lineage of thought gained popular attention and general traction through (among others) Abeba Birhane's seminal paper on the algorithmic colonization in Africa (Birhane, 2020; see also Birhane, 2021).

Accepting AI's false universalism and following Quijano, AI can be said inflect the Eurocentrism that controls intersubjectivity in the modern capitalistic order defined by the coloniality of power on a very large scale (Quijano & Ennis, 2000: p. 545). This necessitates abolishment of epistemological and cognitive barriers instituted and reinforced in the development and deployment of AI systems. Hence the two arguments exposing AI's false universalism correspond to the epistemic reason for decolonizing AI. AI systems risk reifying Western knowledge and values as universal and so reproduce a colonial epistemic dominion. In turn, this links to the ecological reason since abolishment of AI's universalism calls for re-imagining and remodeling the ecological dependencies of creation and usage of AI systems.

### 6.3. Resistance to Precaritisation and Violence

Some authors argue AI systems create precarity among the marginalized and excite several specific types of violence. Their arguments are considered, and they are related to the political and epistemic reasons for decolonization.

McQuillan argues automation and algorithmicizing induce precaritisation of citizens and incite violence. Employing the British Luddites' history of worker resistance to technological development, the Lucas Program's alternative conceptions of technological development (McQuillan, 2023b: p. 8) and case studies of ML-injustices, McQuillan shows AI systems mystify their non-intelligence. The marginalized are increasingly being treated as collateral damage by the solutionist technofixes of AI's abstraction from social problems towards statistical solutions, so McQuillan claims (McQuillan, 2022: p. 45). The idea of AI's enclosing power extends to worker precaritisation and technological support of fascist politics and state power: "AI's solutionism selects some futures while making others impossible to imagine. The question remains as to who's future it will be selecting for" (McQuillan, 2022: p. 45). So how does AI induce precaritisation? Controlled by states and deployed in intra-state contexts, AI is a precaritizing force, because it erodes social elasticity and aggravates (already marginalized) social groups' vulnerability by misuse of corporate/state power. The forces of precaritisation are not exclusive to North or McQuillan's United Kingdom.

Ricaurte depicts its presence in South American governmentality: "the program Technological Platform of Social Intervention of the Argentinian government, who signed a contract with Microsoft to predict teenage pregnancy. Similar cases to surveil children are being developed in Chile and Brazil" (Ricaurte, 2022: p. 735). Precaritisation through AI is thus the weakening by means of automation or algorithmization of social groups' political foundations of welfare and well-being.

How then does AI incite digital violences? Increasingly, AI systems exert algorithmic violence in the service of "states of exception": the state's turn to extra-constitutional brute force to uphold the social status quo, as was seen during the pandemic and the treatment of terrorist threats, in short times. AI can therefore contribute to borderisation of what Santos "the realm of the *lawless*" (Santos, 2014). "The readiness of AI to be applied to borders at all levels," McQuillan explains, "from national territories to cultural and gender norms, can serve to perpetuate violence […]" (McQuillan, 2022: p. 138). Thus, AI yields utility to enclose and keep enclosed. "AI is colonial both because of the intellectual framework it inherits and due to its racialized practices of exteriorization and exclusion" (McQuillan, 2022: p. 136): it places social groups outside of the domain of the law through the exertion of specific forms of violence. McQuillan presents the following typology of (i) *administrative/bureaucratic,* (ii) *epistemic,* and (iii) *hermeneutic* violences (McQuillan, 2022: pp. 53, 60-61). According to McQuillan, these types can be recombined into forms scientistic determinism, behavioral risk indicating systems, racialization, the reproduction of physiognomy and race science, McQuillan claims (McQuillan, 2022: pp. 61, 69-71)[8].

(i) Administrative/bureaucratic violence is the outsourcing of governmental decisions and surveillance actions to AI[9]. These algorithmic decisions become invested with power over citizens' daily affairs, resulting in automated exercise of arbitrary power over life-shaping government-related issues.

(ii) Generally, epistemic violence, is the case when the experience of a social group "only become[s] known through knowledge created by the distant colonial centre" (McQuillan, 2022: p. 62). For example, this is enacted by AI systems when people are falsely categorized, i.e. misgendered or racially profiled (Ricaurte, 2019). As Ricaurte describes: "The narratives produced by [the epistemic violence of] algorithmic mediations are powerful and contribute to the establishment of algorithmic and data imperatives, the imposition of hegemonic algorithmic cultures and, therefore, the establishment of an algorithmic governmentality that sustains a capitalistic, colonial and patriarchal order" (Ricaurte, 2022: p. 732).

---

[8]Pushing back on these ideas, however, is the discussion about the coloniality of data of AI being appropriate ways to term the phenomena of oppression at hand. In this conversation, it is objected that there is a clear discontinuity between historical colonialism and data/AI-colonialism: namely that there is no brutal physical violence present in the latter (Couldry & Mejias, 2019: p. 9). Contrary to the claim that it is the presence or absence of *physical* violence that determines whether there can or there cannot be spoken of continuity, I argue that as previously cited examples of military and police usage of AI show, it is laughable to claim that these technologies somehow operate outside of the colonial sphere of violence because extracting data or outsourcing decision-making in and of itself does not maim people in the process *directly*. If governmental organizations can locate, identify, incarcerate, identify you, deny you the right to speak of disqualify your knowledge, and all these harms can be automatized and made more efficient via AI, grave forms of violence are brought to bear on you that especially reflect the colonial policing, value imposition and racial dehumanization particular of colonial rule.

[9]For example, Israel uses facial recognition to surveil Palestinians and Israelis alike (Amnesty International, 2023). The Netherlands has condemned ethnical minorities to unfair debts and child separations using the SyRi system to indicate potential tax frauds (Rachovitsa & Johann, 2022).

(iii) Lastly, hermeneutic violence—which is a specific type of epistemic injustice—is the disqualification of persons as knowing subjects by AI. A person's interpretation of the world is overruled via the imposition of the "real" that is algorithmically constructed out of data "that must speak for the facts", contra the person's perspective[10].

The identification of precaritisation and bureaucratic, epistemic and hermeneutic violence as characteristic for the politics implicit in and enabled by AI shows the technological continuity between colonialism and AI injustices. This clearly supports the political reason for decolonization, as it holds that as inflection of the historical colonial project, the violences of the AI enabled politics and economy must be decolonized. Likewise, the epistemic reason is constructively emphasized, because the epistemic and hermeneutic forms of violence explain how social groups are harmed through the imposition of Western knowledge and values as universal and the colonization of their subjectivity.

## 6.4. Participation of Affected Social Groups

In this section, a picture of decolonial AI grounded in the participation of affected social groups I sketched. The message the authors work described here conveys is that decolonial AI must be grounded in the world of AI practitioners and place the most marginalized and vulnerable people at the center of AI design and usage. Without practitioners channeling decolonization with help of groups being affected, only colonial AI will come from it, benefitting the existing industry and potentates.

Taking the people directly affected by AI seriously is characteristic of the paradigm of "participatory AI" described by Birhane et al. This paradigm acknowledges "communities and publics beyond technical designers have knowledge, expertise and interests that are essential to AI that aims to strengthen justice and prosperity (Birhane et al., 2022b). Mohamed et al. provide a critical framework stressing the importance of critical theory as *foresight tactics* in AI development. Foresight tactics anticipate "prospective ethical and social harms" (Mohamed, Png, & Isaac, 2020: p. 662) of AI applications and develop new re-search cultures focusing on inclusivity for all stakeholders in AI (Mohamed, Png, & Isaac, 2020: p. 677). On this view, reflexivity on the part of designers and deployers that refuse to forget power's implicit values and inequities are deemed the marks of adequately decolonial AI (Mohamed, Png, & Isaac, 2020: p. 672). Uprooting algorithmic oppression, exploitation, and dispossession is required for decolonial AI. Fostering research in technical areas such as fairness, safety, diversity policy and forms of reciprocal, "reverse" tutelage are needed (Mohamed, Png, & Isaac, 2020: pp. 673-674). Reverse tutelage is the engagement between the colonialized periphery and the colonizing center where the periphery tutors the center on how AI should be developed and used (Mohamed, Png, & Isaac, 2020: p. 674). Effectively, engagement of affected social groups' reverse tutelage with developing parties in the AI economy's sociotechnical foresight translates the political reason for decolonizing AI from theory to practice. Namely because it incorporates the affected in an anticolonial way of developing technology and questions concerning adequate representation of target groups. Correspondingly, possible colonial harms can become recognized in development.

---

[10]Epistemic and hermeneutic violence are combined in the example of the creation of "local Souths" in the global North. The Los Angeles Police Department uses Palantir-manufactured, AI powered surveillance technologies that disproportionally targets Black and Latinx individuals and established a digital form of domination. How? The prior racist hegemony of non-digital surveillance and policing power is reproduced in the software by its being fed mostly data on previous encounters with Black and Latinx individuals that were already the product of ethnic profiling, thus creating a reinforcing loop (Crawford, 2021: pp. 197-198).

Lambrechts et al. study another way for participatory AI to come to fruition. They argue AI needs carefully conducted anthropology of cultures from the global South to be non-replicative of Western values. They propose *value sensitive* algorithm design (VASD) to incorporate the values of specific user groups more completely (Lambrechts, Sinha, & Mosoetsa, 2022). VASD would be capable of apprehending colonizations occurring through the eclipsing social groups' moral and cultural values by dominant Eurocentric ones. This "eclipse" is particular to colonization, e.g. through linguistic oppression (Wiredu, 2002), and resembles the Fanonian/Quijanoan cognitive colonialism. Via target groups' knowledge, participatory AI leads to fertile inflections of sociological studies too, such as the address of the general maltreatment of data production workers in the study of Miceli et al. (2022). Forms of incorporation of target groups' values and knowledges and centralizing non-Western knowledges and values effectively diminish the epistemic dominance of colonial epistemologies otherwise spread by AI.

Apart from these clear promises inherent in participatory AI, Birhane et al. stress the dangers of a naïve adaptation of participation. Firstly, participation is vulnerable to co-optation by industry; participation is unable to replace democratic decision-making; participation isn't equivalent with inclusion; and (there is no "one size fits all" type of participation, as participatory processes tend to reflect the unequal power structures in societies (Birhane et al., 2022b: p. 7). Participatory AI by itself is therefore too context-dependent for a general account of decolonial AI and risks falling prey to co-optation as cover up for continuing injustices. To be specific, "participation" could become a new ethics washing term. Participatory AI can be neutralized before toppling existing oppressive or dominating barriers associated with AI.

Likewise, skepticism is justified with respect to reverse tutelage's similarity to the economic metropole/periphery theories of the mid-twentieth century. Re-verse tutelage risks replicating narratives of Western beneficence and the global North as driver of civilization by virtue of the schematic juxtaposition of a North "allowing" the South to participate *alongside* them (Dussel, 1985). However, the concepts harness deploymental utility, as the periphery identifies the problematic consequences firsthand and so the inclusion of their lived experience in AI design does enable reverse tutelage. The liberating power of the concept reverse tutelage is retained if all Souths enclosed in the North are recognized as well. In turn reverse tutelage should be conceived of as a dialectical process between AI dominator and *any* affected social group, *wherever* situated[11]. This is congruent with the general imperative of constructing an epistemology of the South, at the core of which should be placed the taking serious of experiences and knowledges from the Souths.

## 7. Discussion

AI increasingly pervades human life, from the allocation of governmental freedoms to time spent under AI surveillance and in engagement with algorithmically distributed content. Algorithmically mediated interaction and precaritisation of digital lives is in the process of erecting a novel status quo of enclosure because of the dependency of the digital economy and private sphere on AI-driven, platform companies. The filter bubble that contains preference-tailored, digitally segregated communities is enveloping the world and stands orthogonal to disenclosure because of computationally filtered barriers, borders, and mediations it imposes. Forms of enclosure in service of the existing matrices of domination that must be addressed by decolonizing AI (Muldoon & Wu, 2023).

Having grasped the concepts of colonialism and disenclosure and through exploration and corroboration of the reasons for decolonizing AI, this section discusses the question "what does decolonial AI as disenclosure amount to?"

The argument so far is that the deployment and development of AI is never disconnected from the social sphere; the problematic values AI can spread and the norms it enforces are not accidental (6.3.). AI is never neutral nor innocently scientific, rather AI is deeply rooted in fascist, racist, patriarchal, and colonial superstructures, like Northern societies themselves (6.2.). AI is also completely dependent on an extractivist capture of resources and human life (6.1.). By virtue of the reification of colonialism, the summons of disenclosing of relations to oneself and other beings of decolonial thought (5.) is applicable here.

Using Mbembe's notion of disenclosure, I observe "decolonizing AI" should be about opening AI development and deployment for it to transgress the boundaries of Western universalism. This is the taking down of (preexisting) borders that AI institutes/reinforces within AI design, production, development and deployment. On a larger scale, this is the disenclosure of the entire ecology of AI systems and the ecologies those systems depend on. Mbembe and Fanon saw decolonization as an inherently violent phenomenon of change and decolonizing AI is a summons of only limited value if it fails to be revolutionary (Adams, 2021: p. 1). AI injustices will be reinforced if decolonization is approached as progressive reformations' "new suit". Unlike the historical appropriation of the term decolonization Getachew identified, decolonizing AI is a radical epistemological project. The project needs to revolutionize AI towards something other than service to the nexus of colonialism/digital capitalism. Decolonial epistemology would revolve around a disruption of the intertwinement of the colonialist influences embedded into AI systems (McQuillan, 2022; Katz, 2020) on political, epistemic and ecological fronts. "Opening up" AI development and deployment begins with addressing the particular political, ecological and epistemic barriers in the field.

---

[11]An example to this end is McQuillan's proposal of 'people's councils' of civilians or workers affected by specific AI to provide counterforces against industrialized and governmentalized AI deployment and development (McQuillan, 2022).

The following section discuss ideas generated by the concept of decolonizing AI as disenclosure. First, the political implication of enabling the interconnection of local and global decolonization (7.1.); secondly, the ecological rethinking and reparation of AI ecologies (7.2.); and finally, the empowering of the "wretched of AI" (7.3.).

## 7.1. Political Disenclosure of AI

To address the political reason's global orientation, I claim decolonizing AI has to remain connected the global project of decolonization. Mhlambi and Tiribelli write "The ways to decolonize AI can be as varied as the experiences of colonization are varied. […] Therefore, to create a single framework that adequately addresses the effects of colonization would be infeasible." Accepting this conclusion would undermine the project of this paper. However, we need not to, because the generality of the global decolonial project exists in the desired "common traits" of context-specific local decolonizations that can serve as a way of understanding current and future harms worsened using AI (Mhlambi & Tiribelli, 2023: p. 876).

Politically reconceptualizing AI gathers those commonalities and therefore is, simultaneously, context-sensitive to address the specific histories of the injustices people suffered, and closely connected to general disenclosure that aims to deconstruct colonial oppression on a global scale. This reflects the insight that AI, colonialism and digital capitalism are *global* phenomena that require global responses. Any local or insular attempt at decolonization of AI out of tune with global anticolonial structures is doomed, as failed decolonizations of the twentieth century that relapsed into neoliberal dictatorships attest to[12].

"Decolonial thought is far more than a tool to problematize AI", says Adams (2021: p. 16), because decolonial AI is situated at a crossroads, where either de-tachment or rehabilitation can be chosen. Successful political disenclosure as part of decolonial AI arguably faces the challenge to partake in a politics of rehabilitation. Part of this rehabilitation is to integrate context-sensitive local solutions to AI problematizations into the global project of decolonizing capitalism. Adams also warns that decolonial thought qua reformative tool can be easily co-opted by other forces and directions within the loci of power that are industry and state. A myopic focus on ethics and the participation of marginalized groups in development and deployment that doesn't at the same time address the related extraction of resources constrains decolonization into a social tool with limited application. As part of so-called "decolonial AI", race and historically embedded power structures could risk being merely name-dropped as an ethics-washing facade. Such a departure from the core of decolonial thought must be resisted, as it "is about the production of race and divided worlds; […] it is about knowledge and how knowledge is ascribed legitimacy and value; and it is about a politics of resistance that enters and undoes the object of its critique" (Adams, 2021: p. 16).

[12]Bruno Latour argues with respect to globalization that neither conversative localism nor accelerationist globalism are viable ways to conceptualize the future. Two attractors orthogonal to the global vs. local axis propose two different solutions to the stalemate between the former positions. First, *the out-of-this-world attractor* (which I would explain as the abolishment of the *common world*) and the *Terrestrial attractor* (which I would explain as the rehabilitation of *common world*) (Latour, 2018: p. 40).

"Race and divided worlds in AI systems" are the target of political disenclosure and "knowledge and how knowledge is ascribed legitimacy and value" are the target of epistemic disenclosure (see 7.3.). Decolonial AI in the sense of the political disenclosure of race and divided worlds in AI systems, can be reframed as an engagement in technologies of disenclosure: the making of artifacts and software that abolish boundaries and for which one ensures it obstructs the erection of new oppressive boundaries. Of course, the decolonial capacity of an AI system will depend on the state of (a) the way it is manufactured/designed (who, where, from what resources) and (b) the policies and safeguards in place at the site of its deployment.

How would effectively decolonizing AI influence world economy and development? The authors of the "Decolonial AI Manyfesto" for example call for forms of "decolonial governances" that "will emerge from community and situated contexts, questioning what currently constitutes hegemonic narratives" (Krishnan et al., 2023). Thus, what decolonial AI entails in the governance context is "contextually-relevant" policy, as Gwagwa & Townsend also contend in analyzing the non-applicability of existing regulatory AI frameworks to the African continent (Gwagwa & Townsend, 2023). But in effect, changes in governance such as community involvement and contextual relevance to the global South will form a radical departure from the current status quo of the techno-colonial market, as these anticolonial forms of AI governance problematize the entire colonial supply chain that makes possible the digital capitalism revolving around AI, which touches the heart of the world economy. In the next section, I return to the idea of ecological disenclosure being the redesigning of the ecosystem of extraction AI depends on. Given these ideas, it can be hypothesized that effective decolonization of AI would change the world economy for the better in the sense that decolonial AI governance explicitly targets and con-demns the extractive monopolies in order or local marginalized communities to rise to the occasion to "to decide and build their own dignified socio-technical futures" (Krishnan et al., 2023). If this hypothesis is sensible, then it can also be conjectured that successfully decolonizing AI amounts to taking down the boundaries inherent to the techno-colonial market itself (Mbembe, 2022: p. 67). However, it remains essentially an open question what the global economic landscape will look like after the established of any such politically justified measures.

Anticolonial applications of AI and the politics thereof result from thinking AI and political disenclosure in unison. In sum, it was discussed that it would be beneficial for decolonizing AI technologies to, on the one hand, incorporate local decolonial values to disenclose current value impositions embedded in technology, and, on the other hand, prevent institution of technological enclosures via synchronization with decoloniality on a global level.

## 7.2. Ecological Disenclosure of AI

Along an ecological axis, decolonial AI as disenclosure seems to imply repairing and re-ecologising the planetary systems AI depends on, since AI and computational systems have acquired primacy in digital capitalism's colonial trajectory. Disenclosure, as subclass of reparation, is a foundational countermeasure against colonial instantiations of AI. Decolonizing AI is thus, paradoxically, not only concerned with AI contexts, but also with the extractive ecological weave of colonialism and digital capitalism. Defense of "other sensibilities, cultures and ways of life that do not want to be governed by the market" (Ricaurte, 2022: p. 361) is required and lifting this extractive weight off of particular ecologies requires counterfactual conceptions of their future without AI. Ecological disenclosure, I conjecture, should be understood as a multifaceted reparation of AI ecologies, both global and local.

Breaking down AI's ecological borders requires reclaiming common spaces in the name of those who have been denied access to it, where the "who" is multispecies. The idea of the common finds a powerful metaphorical expression in Mbembe's idea of the capacity for respiration as unifying the living. "The right to breathe" conceived in this way becomes at once a *common that is denied or under threat* for some beings and an *in-common to be shared* by a "we" that is comprised of "all the forces and energies with which we must henceforth learn to live in bio-symbiosis" (Mbembe, 2022: p. 100, 119). The ascendency of the AI economy hides barred ecosystems in its wake. So, for ecological disenclosure, willingness seems to be needed to uproot and redesign the ecologies of knowledge and being that underlie AI and are propagated by it. In line with recent calls to slow the development of LLM's and generative AI in order for regulation and public to catch up, sustainabilization of AI ecologies looks like a prerequisite before any narrative of AI for social good or climate is justified.

Sustainable AI obtained through ecological disenclosure has to see, metaphorically, "from several worlds at a time" (Mbembe, 2022: p. 109), in order not to dominate, deplete or enclose ecologies. Decolonial and disenclosing AI technologies, hailing from Souths and Norths, therefore need to add to a world that is not only post-racial, but also multispecies. An AI development that accounts for the stakes of other species, implies the precondition of AI rehabilitating the ecologies it is premised on. Here again the politics of rehabilitation crosses the trajectory of AI development. Situating AI as a decolonial force therefore means engendering it as a technology of disenclosure that can be used to rehabilitate rather than dominate. In short, to be engaged with entangled re-pairations of AI is first to realize that redistributive social justice for AI depends on ecological reparative justice to provide a viable future for the living. The restorative potential of AI technologies needs to be harnessed to address local as well as global ecological issues, because global social justice already depends on climate justice and the West taking responsibility for the ecocide it has committed (Táíwò, 2022). Rather than automating Western welfare and enclosing Southern societies, I argued AI needs to be developed and deployed in ways that rehabilitate, repair and restore.

### 7.3. Epistemic Disenclosure of AI

At worst, AI is constitutive of "*epistemicide*, the murder of knowledge, by virtue of its false universalism. Unequal exchanges among cultures have always implied the death of the knowledge of the subordinated culture, hence the death of the social groups that possessed it" (Santos, 2014: p. 149). AI designed for the global North simultaneously excludes individuals falling outside of that political, cultural and socioeconomic background. The classificatory, ordering, predictive power of AI becomes the hegemonic cultural nexus of knowledge, eclipsing others' knowledges, which is exactly what the epistemic reasons try to make sensible. I conceive of *epistemic disenclosure* as subspecies of disenclosing the world to all that are denied non-dominated epistemological access to it by *de facto* AI, in the form its administrative, hermeneutic and epistemic violences. In general, this reflects what Ndlovu-Gatsheni problematizes as colonialism's cognitive impact: "taking ideas from a singular "province" of the world and making it into universal knowledge" (Ndlovu-Gatsheni, 2019: p. 222).

Epistemic disenclosure parallels the demystification of AI's non-intelligence. Algorithmic decisions are not statistical eureka's emerging from black boxes, but rather instances in decision- and prediction-making spaces configured by human relations of power. Crawford is similarly occupied with ML-predictions' epistemic status, which she calls "enchanted determinism." "Enchanted" in the sense the obscurity of its workings mystifies, and "determinism" in the sense in which the resulting predictions of AI systems are transfigured into *certainties* (Crawford, 2021: p. 213). From these certainties, narratives of computational interventions as universal solutions and of dystopian heraldry of catastrophe emerge (Crawford, 2021: pp. 214-215). Both should be viewed with suspicion, because they imply intensifications of AI's extractive economy. On the one hand,

"AI promises our salvation, which we must incessantly pursue"; on the other hand, "All research and developmental resources ought to be utilized to stop AI from wreaking havoc among humanity". Both narratives overlap in being (i) long-term views that (ii) intensify capital investments into the digital economy with colonial injustices as collaterals; and (iii) they strive to minimize future suffering while remaining ignorant of actual suffering. Sarcastically speaking, here we have mostly white people from the global North either hailing techno-salvation or sounding the alarm on AI becoming a threat to their capitalistic hegemony. Following Crawford and McQuillan, these can be classified as *falsely justificatory narratives* for the colonial practices, exclusion and harm of affected social groups that sustain AI development and deployment (Couldry & Mejias, 2023: p. 11).

Appropriating Frantz Fanon's dictum, I argue epistemic disenclosure should be about the empowerment of "the wretched of AI" (Fanon, 1963) in the same way that decolonialism is about the restitution of effective freedom to colonialism's "wretched of the Earth" (Ndlovu-Gatsheni, 2019: p. 213), in order to stop the imposition of universalism through AI. The wretched of AI are the social groups marginalized and targeted by AI violences, for example the separated families of the SyRI-system (administrative violence), the discriminated, racially ordered and misrepresented, for example the Black communities that are ethnically profiled (epistemic violence) and those whose voices, values and knowledges are discredited and underrepresented (hermeneutic violence).

Congruent with Adams' reading of decolonial AI, I contend that where AI's racialization, the pursuit of extractive, analytical intelligence, and "manipulation of behavior, attention and thought" are all affirmations of *death*, speaking from the global South's decolonial thought can be "*life-affirming*" and an integral part of the necessary reform. Using Santos' phrase, "…there is no global social justice without global cognitive justice," so epistemological restitutions of effective freedom depend on "*alternative thinking of alternatives*" (Santos, 2014: p. 70). This is reflective of the twentieth-century predicament of fusing feminist/antiracial epistemologies with more general theories of global oppression (such as Marxism) that Charles Mills identified: "a synthesis of these alternative epistemologies, which recognizes both the multiplicity and the unity, the experiential subjectivity and the causal objectivity, of hierarchical class-, gender-, and race-divided society" (Mills, 1998: p. 39). The knowledge production underlying AI systems and digital technologies cannot be left out of this movement of thinking and has to be reconfigured to the wretched of AI. Collective reimaginations need "to be imaginative enough to conceive of a future without AI" (Adams, 2021: p. 17) or at least with AI very different from the kind we know today and to the administrative, epistemic and hermeneutic benefit of the wretched of AI.

## 8. Conclusion

In this paper, the question of conceptualizing decolonial AI was pursued. Sur-veying theories of decolonization, political, ecological and epistemic reasons for why AI systems require decolonization were developed. The concept of disenclosure was presented as suitable conceptualization for decolonization. Using the concept of decolonization as disenclosure, the reasons for decolonizing AI were corroborated by the investigation of four topical currents in decolonial technology studies: critical cartographies of AI systems, arguments exposing AI's false universalism, AI precaritisation and violences and the participation of affected social groups. In conclusion, the conjecture of "decolonial AI as disenclosure" was discussed. The political, ecological and epistemic reasons for decolonial AI as forms of disenclosure opened up new ways to think about and intervene in colonial instantiations of AI development and deployment, to empower "the wretched of AI", re-ecologise the unsustainable ecologies AI depends on and counter the colonial power structures unreflective AI deployment risks to reinforce.

For now, it is up to future work to translate this conceptual approach into lasting worldly changes. For if the use of novel concepts does not enable forms of thinking and intervening in the world, they are to be discarded. The same holds for conceptions of decolonialism and AI that work against those they should aim to free. However, many challenges remain for decolonizing and disenclosing AI. Along a political axis, the discussion already named the danger of co-opting decolonial projects and the complexity of simultaneously fostering local decolonial AI, while remaining connected to global decolonization of digital capitalism's AI. The epistemic project of disenclosing AI seems to face the fiercest opposition from the ethics washing that is constituted by the myopic frameworks of explainable and ethical AI. As long as the questions "Ethical for whom?" and "Explainable to whom?" are left unaddressed, the wretched of AI will keep being epistemically and hermeneutically violated. Finally, from an ecological perspective, decolonizing AI faces many of the same challenges that have kept global efforts to address climate change from coming off the ground: changing the material underpinnings of the AI economy is not in the direct interest of the network of power driving its acceleration. In an ever-developing world the "*remaking of the world*" needs to be recognized as "permanent activity" (Mbembe, 2022: p. 124) that proceeds via "rethinking our imperial past and present in the service of imagining an anti-imperial future" (Getachew, 2019: p. 181). With that in mind the biggest challenge for decolonial AI emerges: to convince those shaping AI development of the shared duty "to our moral descendants" (Táíwò, 2022: p. 207) of prohibiting AI deployment to continue along its colonial trajectory.

## Acknowledgements

The author would like to thank Arthur Gwagwa for his helpful comments on earlier drafts of this paper. Thanks also go out to Aranea Ham, for her comments and her continuous availability for discussing the ideas developed in this paper.

## Conflicts of Interest

The author declares no conflicts of interest regarding the publication of this paper.